\pdfoutput=1

\documentclass[10pt,conference]{IEEEtran}

\usepackage{cite}
\usepackage{amsmath,amssymb,amsfonts}   
\usepackage{graphicx}
\usepackage{textcomp}
\usepackage{xcolor}
\usepackage[caption=false,font=footnotesize]{subfig}
\usepackage{stfloats}
\usepackage{epstopdf}
\usepackage{acronym}
\usepackage{mathrsfs}
\usepackage{placeins}
\usepackage{flushend}
\usepackage{booktabs}
\usepackage{multicol}
\usepackage{hyperref}
\usepackage{float}
\usepackage{svg}

\def\BibTeX{{\rm B\kern-.05em{\sc i\kern-.025em b}\kern-.08em
		T\kern-.1667em\lower.7ex\hbox{E}\kern-.125emX}}
        
\acrodef{WSN}{wireless sensor network}
\acrodef{USRP}{universal software radio peripheral}
\acrodef{SN}{sensor node}
\acrodef{FC}{fusion center}
\acrodef{MAC}{multiple-access channel}
\acrodef{FL}{federated learning}
\acrodef{ED}{edge device}
\acrodef{CS}{compressed sensing}
\acrodef{ES}[BS]{base station}
\acrodef{DCN}{data center network}
\acrodef{RIS}{reconfigurable intelligent surfaces}
\acrodef{IMC}{in-memory computing}
\acrodef{FPGA}{field-programmable gate array}
\acrodef{SDR}{software-defined radio}
\acrodef{PS}{processing system}
\acrodef{SS}{soft synchronization}
\acrodef{IQ}{in-phase/quadrature}
\acrodef{IP}{intellectual property}
\acrodef{DMA}{direct-memory access}
\acrodef{RAM}{random access memory}
\acrodef{CC}{companion computer}
\acrodef{FEE}{function estimation error}
\acrodef{MSK}{minimum-shift keying}
\acrodef{TDMA}{time-domain multiple access}
\acrodef{PLNC}{physical-layer network coding}
\acrodef{UAV}{unmanned aerial vehicle}
\acrodef{LoRa}{Long-Range}
\acrodef{DC}{direct-current}
\acrodef{DAC}{digital-to-analog converter}
\acrodef{ADC}{anlog-to-digital converter}
\acrodef{CS}{complementary sequence}
\acrodef{GCP}{Golay complementary pair}
\acrodef{ANF}{algebraic normal form}
\acrodef{AACF}{aperiodic auto-correlation function}
\acrodef{AACFs}{aperiodic auto-correlation functions}
\acrodef{RM}{Reed-Muller}
\acrodef{MOCZ}{modulation on conjugate-reciprocal zeros}
\acrodef{BMOCZ}{binary MOCZ}
\acrodef{dizet}[DiZeT]{direct zero-testing}

\acrodef{NN}{neural network}
\acrodef{PUCCH}{physical uplink control channel}
\acrodef{PRACH}{physical random access channel}

\acrodef{OBO}{output-power back-off}
\acrodef{ACLR}{adjacent-channel-leakage ratio}

\acrodef{LDPC}{low-density parity check}

\acrodef{PDF}{probability density function}
\acrodef{CDF}{cumulative distribution function}
\acrodef{CCDF}{complementary cumulative distribution function}

\acrodef{TBMA}{type-based multiple access}

\acrodef{MSFE}{mean-squared function error}
\acrodef{FEE}{function-estimation error}
\acrodef{CER}{computation error rate}
\acrodef{BCER}{block-computation error rate}
\acrodef{CFO}{carrier frequency offset}
\acrodef{TO}{time offset}
\acrodef{PO}{phase offset}
\acrodef{RSSI}{received signal strength  information}

\acrodef{STLC}{space-time line code}
\acrodef{CCI}{co-channel interference}
\acrodef{CSIT}[CSIT]{\ac{CSI} at the transmitter}
\acrodef{CSIR}[CSIR]{\ac{CSI} at the receiver}
\acrodef{MIMO}{multiple-input multiple-output}
\acrodef{PC}{phase correction}
\acrodef{ZF}{zero-forcing}
\acrodef{ANOVA}{analysis of variance}

\acrodef{PCA}{principal component analysis}
\acrodef{TIG}{Technical Interest Group}

\acrodef{FSK}{frequency-shift keying}
\acrodef{PPM}{pulse-position modulation}
\acrodef{PAM}{pulse-amplitude modulation}

\acrodef{MRC}{maximum-ratio combining}
\acrodef{HP}{hard-coded participation}
\acrodef{HPA}{hard-coded participation with absentees}
\acrodef{SP}{soft-coded participation}
\acrodef{FSK-MV}{\ac{FSK}-based \ac{MV}}
\acrodef{RF}{radio-frequency}
\acrodef{MF}{matched filter}
\acrodef{PPM}{pulse-position modulation}
\acrodef{CSK}{chirp-shift keying}
\acrodef{PPM-MV}[PPM-MV]{\ac{PPM}-based \ac{MV}}
\acrodef{DFT-s-OFDM}{discrete Fourier transform-spread orthogonal frequency division multiplexing}
\acrodef{SC}{single-carrier}
\acrodef{SGD}{stochastic gradient descent}
\acrodef{signSGD}{sign stochastic gradient descent}

\acrodef{SL}{split learning}
\acrodef{SNR}{signal-to-noise ratio}
\acrodef{RMSE}{root-mean-squared error}
\acrodef{OFDM}{orthogonal frequency division multiplexing}
\acrodef{DFT}{discrete Fourier transform}
\acrodef{PSK}{phase-shift keying}
\acrodef{QAM}{quadrature amplitude modulation}
\acrodef{QPSK}{quadrature phase-shift keying}
\acrodef{PMEPR}{peak-to-mean envelope power ratio}
\acrodef{BER}{bit error rate}
\acrodef{SNR}{signal-to-noise ratio}
\acrodef{PSD}{power spectral density}
\acrodef{SE}{spectral efficiency}
\acrodef{CP}{cyclic prefix}
\acrodef{AWGN}{additive white Gaussian noise}
\acrodef{CFR}{channel frequency response}
\acrodef{CIR}{channel impulse response}
\acrodef{MMSE}{minimum mean-squared error}
\acrodef{LMMSE}{linear minimum mean-squared error}
\acrodef{BPSK}{binary phase shift keying}
\acrodef{BPSK}{quadrature phase shift keying}
\acrodef{BLER}{block error rate}
\acrodef{PHY}{physical layer}
\acrodef{PA}{power amplifier}
\acrodef{IDFT}{inverse discrete Fourier transform}
\acrodef{DoF}{degrees-of-freedom}
\acrodef{IoT}{Internet of Things}
\acrodef{mMTC}{massive machine-type communication}
\acrodef{URLLC}{ultra-reliable low-latency communication}
\acrodef{FDE}{frequency-domain equalization}
\acrodef{RF}{radio-frequency}
\acrodef{IM}{index modulation}
\acrodef{MF}{matched filter}
\acrodef{PPM}{pulse-position modulation}

\acrodef{MSE}{mean-squared error}
\acrodef{MRT}{maximum-ratio transmission}
\acrodef{ERC}{equal-ratio combining}
\acrodef{BAA}{broadband analog aggregation}
\acrodef{OBDA}{one-bit broadband digital aggregation}
\acrodef{FEEL}{federated edge learning}
\acrodef{FL}{federated learning}
\acrodef{UL}{uplink}
\acrodef{OAC}{over-the-air computation}
\acrodef{TCI}{truncated-channel inversion}
\acrodef{MV}{majority vote}
\acrodef{CNN}{convolution neural network}
\acrodef{ReLU}{rectified-linear unit}
\acrodef{CSI}{channel state information}
\acrodef{PAPR}{peak-to-average power ratio}
\acrodef{SC}{single-carrier}
\acrodef{iid}[IID]{independent and identically distributed}
\acrodef{RMS}{root-mean-square}
\acrodef{4G}{fourth generation}
\acrodef{5G}{Fifth Generation}
\acrodef{6G}{Sixth Generation}
\acrodef{NR}{New Radio}
\acrodef{LTE}{Long-Term Evolution}
\acrodef{OFDMA}{orthogonal frequency division multiple access}
\acrodef{HARQ}{hybrid automatic repeat request}
\acrodef{D2D}{Device-to-Device}
\acrodef{NOMA}{non-orthogonal multiple access}
\acrodef{OMA}{orthogonal multiple access}
\acrodef{IMT}{International Mobile Telecommunications}
\acrodef{ITU}{International Telecommunication Union}

\acrodef{PDP}{power-delay profile}
\acrodef{TBMA}{type-based multiple access}
\acrodef{ISI}{intersymbol interference}
\acrodef{MLSE}{maximum likelihood sequence estimator}
\acrodef{LTI}{linear time-invariant}
\acrodef{ISAC}{integrated sensing and communication}
\acrodef{ML}{machine learning}
\acrodef{DL}{deep learning}
\acrodef{AE}{autoencoder}
\acrodef{AEs}{autoencoders}
\acrodef{MLP}{multi-layer perceptron}
\newcommand{\trans}{^\text{T}}
\newcommand{\herm}{^\text{H}}
\newcommand{\vecsym}[1]{\boldsymbol{\rm{#1}}}

\def\vector{\vecsym{v}}

\def\complexnormal{\mathcal{CN}}

\def\complexnumbers{\mathbb{C}}
\def\realnumbers{\mathbb{R}}
\def\zeroCodebook{\mathscr{Z}}

\def\radius{R}
\def\numZeros{K}
\def\zeroIndex{k}
\def\zero{\alpha}

\def\messageSeq{\vecsym{b}}
\def\polySeqTX{\vecsym{x}}
\def\polySeqRX{\vecsym{y}}

\def\dizetRadius{\radius_\text{DZ}}

\def\totLength{L_\mathrm{t}}

\def\zeroVector{\boldsymbol{\alpha}}

\def\filter{\vecsym{h}}
\def\numTaps{L}
\def\noiseSeq{\vecsym{w}}
\def\estMessageBit{\hat{b}}

\def\bzerosestimate{\hat{\boldsymbol{\alpha}}}
\def\zerosestimate{\hat{\alpha}}
\def\realcomponent[#1]{\text{Re}\left(#1\right)}
\def\imaginarycomponent[#1]{\text{Im}\left(#1\right)}
\def\companionmatrix{\vecsym{C}}
\def\mlpParameters{\boldsymbol{\theta}_{\rm NN}}
\def\logodds{p}
\def\blogodds{\mathbf{p}}

\def\tmessage{\tilde{b}}

\def\numMessages{B}
\def\messageIndex{j}

\def\phaseParameters{\boldsymbol{\theta}_{\rm c}}
\def\xMLP{\mathbf{x}_\text{NN}}
\def\complexity{\mathcal{O}}

\def\ebno{E_{\rm b}/N_0}
\def\hiddenlayers{L_{\rm hidden}}
\def\dvariable{\tau}

\def\threshold{t}
\def\lparametersDZ{\mathcal{P}_\text{DZ}}
\def\lparametersMLP{\mathcal{P}_\text{NN}}
\def\jthExample{^{(j)}}
\def\lossa{\mathcal{L}}
\def\lossb{\mathfrak{L}}

\def\e{\mathrm{e}}
\def\X{X}
\def\H{H}
\def\W{W}
\def\Y{Y}

\begin{document}
	
    \title{Learning Zero Constellations for Binary MOCZ in Fading Channels}
	
    \author{Anthony Joseph Perre, Parker Huggins, and Alphan \c{S}ahin\\
    Department of Electrical Engineering, University of South Carolina, Columbia, SC, USA\\
    Email: \{aperre, parkerkh\}@email.sc.edu, asahin@mailbox.sc.edu 
    }
	
    \maketitle
    
    \begin{abstract}  	
        In this work, we propose two methods to design zero constellations for binary modulation on conjugate-reciprocal zeros (BMOCZ). In the first approach, we treat constellation design as a multi-label binary classification problem and learn the zero locations for a \ac{dizet} decoder. In the second approach, we introduce a \ac{NN}-based decoder and jointly learn the decoder and zero constellation parameters. We show that the \ac{NN}-based decoder can directly generalize to flat-fading channels, despite being trained under additive white Gaussian noise. Furthermore, the results of numerical simulations demonstrate that learned zero constellations outperform the canonical, Huffman BMOCZ constellation, with the proposed \ac{NN}-based decoder achieving large performance gain at the expense of increased computational complexity.  
    \end{abstract}
    
    
	\acresetall
    
    \section{Introduction} \label{sec:intro}

    Non-coherent communication offers a promising solution to reduce overhead in modern wireless systems. Since the energy of the received symbols are utilized for signal detection, the receiver in a non-coherent communication system does not require the \ac{CSI}. This characteristic eliminates the need for extensive overhead signaling, resulting in increased bandwidth efficiency and large energy savings, particularly for \ac{MIMO} systems~\cite{xu2019sixty}. Despite these advantages, non-coherent schemes suffer from degraded performance. Therefore, further research is required to develop non-coherent solutions that are both reliable and practical~\cite{chafii2023twelve}.
    
    A recently proposed non-coherent communication scheme is \ac{MOCZ}~\cite{walk2017short}. The principle of \ac{MOCZ} is to encode information into the zeros of the baseband signal's $z$-transform. With this approach, the receiver does not require knowledge of the \ac{CIR}, since the zeros of the transmitted polynomial are preserved after passing through the channel. In~\cite{walk2019principles} and~\cite{walk2020practical}, the authors introduce a \ac{BMOCZ} scheme, called Huffman \ac{BMOCZ}, together with a \ac{dizet} decoder. For Huffman \ac{BMOCZ}, the coefficients of the transmitted polynomial form a Huffman sequence, which have an impulse-like auto-correlation function~\cite{ackroyd1970design}. There are several studies exploring the wide-ranging applications of Huffman \ac{BMOCZ}, including works on multi-user access~\cite{walk2021multi}, integrated sensing and communication~\cite{dehkordi2023integrated}, and over-the-air computation~\cite{csahin2024over}.   

    Recently, \ac{ML} algorithms have been applied to develop solutions for the physical layer. For example, the authors in~\cite{o2017introduction} interpret communication as a \ac{DL} problem and demonstrate the applicability of various network architectures for transceiver design and modulation classification. Additionally, \ac{DL}-based \ac{AEs} are developed for \ac{OFDM} and multi-user systems in \cite{felix2018ofdm} and \cite{wu2020deep}, respectively. To the knowledge of the authors, however, ML-based solutions for \ac{BMOCZ} have not been investigated in the literature. In fact, excluding the design of alternative \ac{MOCZ} codebooks for \ac{PAPR} reduction in~\cite{sasidharan2024alternative}, there are no works discussing zero constellation design beyond Huffman \ac{BMOCZ}. 
    
    Hence, in this study, we propose two \ac{ML} methods to design zero constellations for \ac{BMOCZ}. Our main contributions are as follows:
    
    \begin{itemize}
    	\item We introduce a continuous-output implementation of the \ac{dizet} decoder, originally proposed in~\cite{walk2019principles}, and use it to learn \ac{BMOCZ} zero constellations via a cost function.
    	\item We propose a \ac{NN}-based decoder that takes the zeros of a received polynomial as inputs and outputs the detected bits. During training, we jointly learn the \ac{NN} and zero constellation parameters.
    	\item We show that the proposed \ac{NN}-based decoder trained in \ac{AWGN} generalizes to fading channels without additional inputs or \ac{CSI}.
    \end{itemize}
    

    \emph{Organization}: The paper is organized as follows: Section~\ref{sec:model} reviews \ac{BMOCZ}, the \ac{dizet} decoder, and constellation design. Section~\ref{sec:methods} describes the proposed \ac{ML} framework, including the learning of zero constellations, \ac{NN}-based decoder design, and complexity analysis. Section~\ref{sec:results} presents the learned zero constellations and the results of simulations in both \ac{AWGN} and flat-fading channels. Finally, we conclude in Section~\ref{sec:conclusion}.

    \emph{Notation}: The sets of real and complex numbers are denoted by $\realnumbers$ and $\complexnumbers$, respectively. The complex conjugate of $z\in\complexnumbers$ is expressed as $z^\ast$, and the real and imaginary parts of $z$ are denoted by $\Re(z)$ and $\Im(z)$, respectively. We denote the Euclidean norm of a vector $\vector\in\complexnumbers^{N}$ as $||\vector||_2=\sqrt{\vector\herm\vector}$. The complex normal distribution with mean zero and variance $\sigma^2$ is denoted by $\complexnormal(0,\sigma^2)$. By $\left[N\right]=\{0,1,\dots,N-1\}$ we denote the set of the first $N$ non-negative integers. For training, the instance of a variable $x$ in the $i$th batch is denoted by $x^{(i)}$.
  
    \section{System model} \label{sec:model}

    Consider a message $\messageSeq=[b_0,b_1,\hdots,b_{\numZeros-1}]\in\{0,1\}^\numZeros$. For Huffman \ac{BMOCZ}, the $\zeroIndex$th message bit is mapped to the $\zeroIndex$th zero of a polynomial according to
    \begin{equation} \label{eq:zero_mapping}
        \zero_k = 
        \begin{cases}
            \radius \, \e^{j2\pi\frac{\zeroIndex}{\numZeros}}, & b_\zeroIndex=1\\
            \radius^{-1} \, \e^{j2\pi\frac{\zeroIndex}{\numZeros}}, & b_\zeroIndex=0
        \end{cases}\:, \ \ \zeroIndex\in[\numZeros]\:,
    \end{equation}
    where $\radius>1$. With~\eqref{eq:zero_mapping}, each message $\messageSeq\in\{0,1\}^\numZeros$ maps to a distinct \emph{zero pattern} $\zeroVector=[\zero_0,\zero_1,\hdots,\zero_{\numZeros-1}]\in\complexnumbers^\numZeros$, which uniquely defines the $\numZeros$th-degree polynomial
    \begin{equation} \label{eq:polynomial}
        \X(z)=\sum_{k=0}^{\numZeros}x_\zeroIndex z^\zeroIndex=x_\numZeros \prod_{\zeroIndex=0}^{\numZeros-1}(z-\zero_\zeroIndex)\:,
    \end{equation}
    up to the factor $x_\numZeros\neq0$. We call $\X(z)$ a \emph{Huffman polynomial}, for its $\numZeros+1$ coefficients $\polySeqTX=[x_0,x_1,\hdots,x_K]\in\complexnumbers^{\numZeros+1}$ form a Huffman sequence~\cite{ackroyd1970design,walk2017short}. In the context of wireless communications, Huffman sequences have several desirable characteristics, including near-ideal \ac{AACFs}. Additionally, the authors in~\cite{walk2019principles} show that the zeros of Huffman polynomials remain stable under additive noise perturbing their coefficients. 

    Assuming a \ac{LTI} channel and invoking the convolution theorem with $\totLength=\numZeros+\numTaps$, the received sequence $\polySeqRX=[y_0,y_1,\hdots,y_{\totLength-1}]\trans\in\complexnumbers^\totLength$ can be expressed in the $z$-domain as 
    \begin{equation} \label{eq:z_domain_mult}
        \Y(z)=\X(z)\H(z)+\W(z)\:,
    \end{equation}
    where $\H(z)$ and $\W(z)$ are the $z$-domain representations of the $\numTaps$-tap \ac{CIR} $\filter=[h_0,h_1,\hdots,h_{\numTaps-1}]\in\complexnumbers^\numTaps$ and the noise sequence $\noiseSeq=[w_0,w_1,\hdots,w_{\totLength-1}]\in\complexnumbers^{\totLength}$, respectively. 
    Hence, the received polynomial $\Y(z)$ has $\totLength-1$ zeros, comprising the $\numZeros$ data zeros $\{\zero_\zeroIndex\}$ and the $\numTaps-1$ channel zeros. At the receiver, an estimate of the transmitted message bits can be obtained using a simple \ac{dizet} decoder~\cite{walk2019principles}. The \ac{dizet} decoder works by evaluating the received polynomial at the zeros in each conjugate-reciprocal zero pair $\zeroCodebook_k=\{\zero_k,1/\zero_k^\ast\}$. The $\zeroIndex$th message bit is then estimated as
    \begin{equation} \label{eq:dizet_decoder}
        \estMessageBit_k = 
        \begin{cases}
            1, & |\Y(\radius \, \e^{j2\pi\frac{\zeroIndex}{\numZeros}})| < \radius^{\totLength-1}|\Y(\radius^{-1} \, \e^{j2\pi\frac{\zeroIndex}{\numZeros}})|\\
            0, & \text{otherwise}
        \end{cases}\:.
    \end{equation}
    Note that the additional $\numTaps-1$ channel zeros can degrade the \ac{BER} performance of the \ac{dizet} decoder. In this study, however, we consider \ac{BMOCZ} within an \ac{OFDM} framework. By appropriately mapping the polynomials to time-frequency resources, the effective channel length can be constrained to unity within the coherence time and bandwidth of the channel~\cite{huggins2024optimal}. Consequently, no zeros are introduced to the transmitted polynomials, i.e., $\totLength=\numZeros+1$. 
    
    While simple, the performance of the \ac{dizet} is sensitive to the \emph{zero constellation} $\{\zero_0,1/\zero_0^\ast,\hdots,\zero_{\numZeros-1},1/\zero_{\numZeros-1}^\ast\}$, since the placement of the zeros in the complex plane determines their stability under noise. For this reason, the optimal choice for the radius in~\eqref{eq:zero_mapping} has been investigated in various works; see, for example,~\cite{walk2019principles} and~\cite{huggins2024optimal}. The current convention for Huffman \ac{BMOCZ} with the \ac{dizet} decoder is to choose the radius $\radius$ that maximizes the pairwise separation of the zeros. The expression for such a radius was identified in~\cite{walk2019principles} as
    \begin{equation} \label{eq:dizet_radius}
    	\dizetRadius(\numZeros,\lambda) = \sqrt{1 + 2\lambda\sin(\pi/\numZeros)}\:,
    \end{equation}
    where $\lambda$ is a weighting parameter that controls the trade-off between the zero separation in radial and angular directions. While setting $\lambda=1$ maximizes the pairwise zero separation, $\lambda\approx1/2$ was heuristically identified in~\cite{walk2019principles} through \ac{BER} simulations and polynomial perturbation analysis. However, to the best of our knowledge, the encoding in~\eqref{eq:zero_mapping} that minimizes the \ac{BER} for \ac{BMOCZ} is unknown in the literature. 
    
    \section{Methodology} \label{sec:methods}

\begin{figure*}[t]
	\centering
        \subfloat[\ac{dizet}-based scheme with learned zero mapper.]
        {\includegraphics[width=6in]{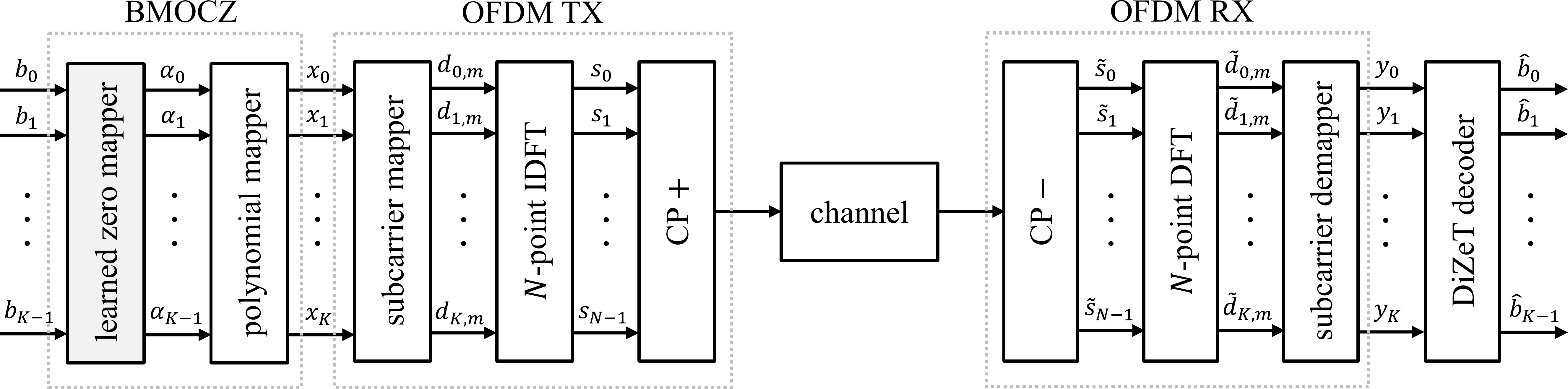}}\label{subfig:dizet_diagram}
        \vspace{2pt}
        \subfloat[\ac{NN}-based scheme with learned zero mapper.]
        {\includegraphics[width=\textwidth]{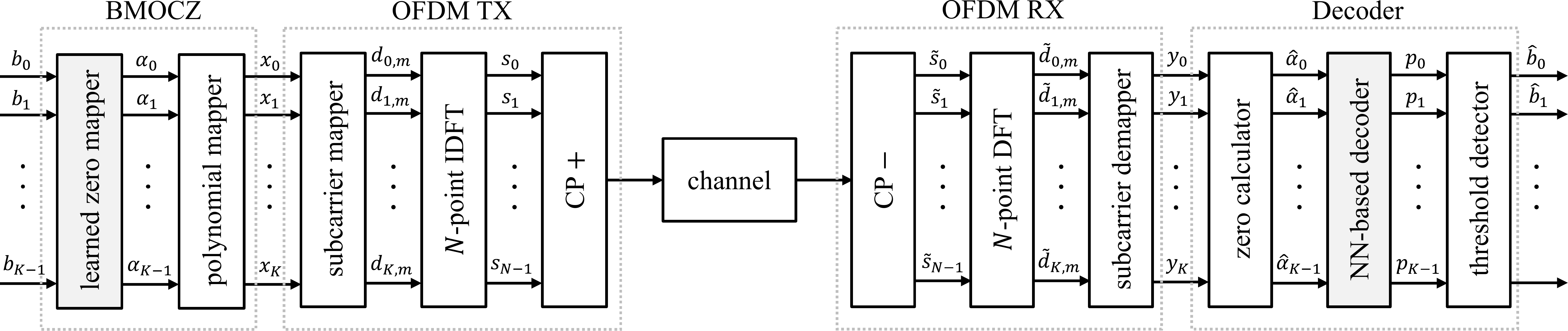}}\label{subfig:mlp_diagram}
        \caption{Processing of the $m$th \ac{OFDM} symbol for \ac{BMOCZ}-based \ac{OFDM} implemented with \ac{dizet} and \ac{NN}-based decoders. For reference, the system blocks learned in this work have been highlighted.}	
        \label{fig:block_diagrams}
    \end{figure*}

    In this study, we consider a single-link \ac{BMOCZ}-based \ac{OFDM} system implemented with frequency-mapping~\cite{huggins2024optimal}, as illustrated in Fig.~\ref{fig:block_diagrams}. The following subsections describe how we optimize the corresponding zero constellation parameters for the \ac{dizet} decoder and an \ac{NN}-based decoder.

    \subsection{DiZeT Zero Constellation Design} \label{subsec:enc}

    Let $\lparametersDZ = \{\radius,\phaseParameters\}$ denote the constellation parameters of a \ac{BMOCZ} scheme, where $\phaseParameters=[\theta_0,\theta_1,\hdots,\theta_{\numZeros-1}]\in[0,2\pi)^\numZeros$ is a vector containing the phase for each possible zero. To learn a zero constellation suitable for the \ac{dizet} decoder, we optimize $\lparametersDZ$ to improve \ac{BER} performance. Assume that we transmit $\numMessages$ messages, each containing $\numZeros$ bits. For the $\zeroIndex$th zero pair tested on the $\messageIndex$th received polynomial $\Y\jthExample(z)$, we define the decision variable
    \begin{equation}
    \dvariable_{\zeroIndex}\jthExample=|\Y\jthExample(\radius \, \e^{j\theta_\zeroIndex})| - \radius^{\totLength-1}|\Y\jthExample(\radius^{-1} \, \e^{j\theta_\zeroIndex})|\:,
    \end{equation} 
    for $\messageIndex\in[\numMessages]$ and $\zeroIndex\in[\numZeros]$. This expression is analogous to~\eqref{eq:dizet_decoder}, where the cases $\dvariable_{\zeroIndex}\jthExample<0$ and $\dvariable_{\zeroIndex}\jthExample\geq0$ correspond to $\estMessageBit_{\zeroIndex}\jthExample=1$ and $\estMessageBit_{\zeroIndex}\jthExample=0$, respectively. By defining $\dvariable_{\zeroIndex}\jthExample$ in this manner, we model the output of the \ac{dizet} decoder as a continuous variable, where the magnitude $|\dvariable_{\zeroIndex}\jthExample|$ reflects the confidence for each predicted $\estMessageBit_{\zeroIndex}\jthExample$. We remark that the use of a continuous output helps the stability of gradient descent methods during training.

    To minimize \ac{BER}, we seek to maximize the margin between correctly and incorrectly classified bits. Hence, we utilize hinge loss~\cite{tsvieli2023learning}, which penalizes both incorrectly classified bits and correctly classified bits for $|\dvariable_{\zeroIndex}\jthExample|<\threshold$, where $\threshold$ is a hyperparameter that controls the margin threshold. For the $\zeroIndex$th bit decoded from the $\messageIndex$th received polynomial, we define the loss function as
    \begin{equation} \label{eq:loss_dizet}
        \begin{aligned}
         \lossa_{\zeroIndex}\jthExample = \max\{0,\threshold-\tmessage_{\zeroIndex}\jthExample \dvariable_{\zeroIndex}\jthExample\}\:,
        \end{aligned}
    \end{equation} 
    where the label $\tmessage_{\zeroIndex}\jthExample \in\{-1,1\}$ is calculated using the affine transformation $\tmessage_{\zeroIndex}\jthExample = 2b_{\zeroIndex}\jthExample - 1$. We then define the optimization problem as
    \begin{equation}
        \lparametersDZ=\arg \min_{\radius,\phaseParameters}\frac{1}{\numMessages\numZeros}\sum_{\messageIndex=0}^{\numMessages-1}\sum_{\zeroIndex=0}^{\numZeros-1} \lossa_{\zeroIndex}\jthExample\:.
    \end{equation} 
    It is important to note that the loss function in \eqref{eq:loss_dizet} differs from the one used in the \ac{NN}-based decoding scheme, as discussed in Section \ref{subsec:dec}.
    
    \subsection{Neural Network Decoder} \label{subsec:dec}
    
    Let $\delta(\xMLP)$ denote an \ac{NN} which maps the input $\mathbf{x}_{\rm NN}$ to a vector $\blogodds=[\logodds_0,\logodds_1,\dots,\logodds_{\numZeros-1}] \in \realnumbers^{\numZeros}$, which represents the logarithmic odds for each bit in $\messageSeq$. In this work, we use the zero estimates $\bzerosestimate = [\zerosestimate_0, \zerosestimate_1,\dots,\zerosestimate_{\numZeros-1}] \in \complexnumbers^{\numZeros}$ as input to the \ac{NN} instead of polynomial coefficients. The rationale behind this choice is discussed in Section~\ref{sec:general}. Let $\Y\jthExample(z)$ be the $\messageIndex$th received polynomial with coefficients $\polySeqRX\jthExample$. To obtain the zero locations of $\Y\jthExample(z)$, we utilize a companion matrix-based approach.\footnote{This method is elected for its compatibility with PyTorch's automatic differentiation engine.} First, we define the Frobenius companion matrix of $\Y\jthExample(z)$ as
    \begin{equation}
        \begin{aligned} \label{eq:companion_matrix}
            \companionmatrix\jthExample &=
            \begin{bmatrix}
                0 & 0 & \cdots & 0 &- {y_{0}\jthExample} / {y_{\numZeros}\jthExample} \\
                1 & 0 & \cdots & 0 &- {y_{1}\jthExample} / {y_{\numZeros}\jthExample} \\
                0 & 1 & \cdots & 0 &- {y_{2}\jthExample} / {y_{\numZeros}\jthExample} \\
                \vdots & \vdots & \ddots & \vdots & \vdots \\
                0 & 0 & \cdots & 1 &- {y_{\numZeros-1}\jthExample} / {y_{\numZeros}\jthExample} 
            \end{bmatrix}\in\complexnumbers^{\numZeros\times\numZeros}\:,
        \end{aligned}
    \end{equation} 
    where the eigenvalues of $\companionmatrix\jthExample$ are the roots of $\Y\jthExample(z)$, i.e., the zeros in $\bzerosestimate\jthExample$. Since a neural network has real-valued inputs, we define a bijection $f : \complexnumbers^\numZeros \to \realnumbers^{2\numZeros}$ that maps each $\bzerosestimate\jthExample$ to a real-valued vector as
    \begin{equation} \label{eq:realmapper}
        \begin{aligned}
            f(\bzerosestimate\jthExample) = [\Re(\zerosestimate_{0}\jthExample), \Im(\zerosestimate_{0}\jthExample),\dots,\Re(\zerosestimate_{\numZeros-1}\jthExample), \Im(\zerosestimate_{\numZeros-1}\jthExample)]\:.
        \end{aligned}
    \end{equation} 
    Note that the roots estimated from the companion matrix in~\eqref{eq:companion_matrix} are not returned in any particular order. Hence, the \ac{NN} must remain permutation agnostic with respect to the real-imaginary zero pairs in \eqref{eq:realmapper}, which increases the overall model complexity~\cite{kimura2024permutation}. Letting $\blogodds\jthExample = (\delta\circ f)(\bzerosestimate\jthExample)$, the decoding rule for the $\zeroIndex$th bit obtained from $\Y\jthExample(z)$ becomes a hard decision on $\logodds_\zeroIndex\jthExample$ as 
    \begin{equation} \label{eq:mlp_decoder}
        \estMessageBit_{\zeroIndex}\jthExample = 
        \begin{cases}
            1, & \logodds_\zeroIndex\jthExample > 0\\
            0, & \text{otherwise}
        \end{cases}\:.
    \end{equation}
    
    For this approach, the learnable parameters are defined as $\lparametersMLP=\{\radius,\phaseParameters,\mlpParameters\}$, where $\mlpParameters$ represents the weights and biases in the \ac{NN}. To optimize $\lparametersMLP$ in terms of \ac{BER} performance, we attempt to minimize the binary cross-entropy loss given by
    \begin{equation} \label{eq:loss2}
        \begin{aligned}
            \lossb_{\zeroIndex}\jthExample &= -b_{\zeroIndex}\jthExample\ln(\sigma(\logodds_{\zeroIndex}\jthExample))-(1-b_{\zeroIndex}\jthExample)\ln(1-\sigma(\logodds_{\zeroIndex}\jthExample))\:,
        \end{aligned}
    \end{equation} where $\sigma(\cdot)$ denotes the sigmoid function. For $B$ transmitted polynomials, the optimization problem is then defined as
    \begin{equation}
        \lparametersMLP=\arg \min_{\radius,\phaseParameters,\mlpParameters}\frac{1}{B\numZeros}\sum_{\messageIndex=0}^{\numMessages-1}\sum_{\zeroIndex=0}^{\numZeros-1} \lossb_{\zeroIndex}\jthExample\:.
    \end{equation} 
     In this scheme, the zero constellation parameters $\{\radius,\phaseParameters\}$ and the \ac{NN}-based decoder parameters $\mlpParameters$ are jointly learned using ADAM. Furthermore, a two-step training process is implemented to ensure robust performance across a wide range of \ac{SNR}, which is discussed in Section~\ref{sec:train}.
     
    \subsection{Model Generalization} \label{sec:general}
    The motivation for using the zeros instead of the received polynomial coefficients for detection is to eliminate the need for \ac{CSI}. For example, in a flat-fading channel, the polynomial $\H(z)$ reduces to a constant, i.e., $\H(z)=h$ for some $h\in\complexnumbers$. Under this condition, we can express the received polynomial in \eqref{eq:z_domain_mult} as 
      \begin{equation} \label{eq:z_domain_mult2}
     	\Y(z)= h\left(\X(z)+\frac{W(z)}{h}\right)\:.
     \end{equation} 
     Observe that the scalar $h$ does not alter the distribution of zeros for $\W(z)$. Instead, $h$ simply scales the energy of $\X(z)$ and $\W(z)$, thereby controlling the extent to which $\W(z)$ perturbs the roots of $\X(z)$. Hence, a model trained under \ac{AWGN} can directly generalize to flat-fading channels, a unique advantage of \ac{BMOCZ} over traditional modulations, since the distribution of the zeros of $\W(z)$ is the same for both cases. Moreover, training in an \ac{AWGN} channel instead of a fading channel offers the benefit of reduced variance in the training samples, which helps the \ac{NN} to learn the global structure of each polynomial.
     
    \subsection{Training Strategy} \label{sec:train}
    First, we consider the scheme with the \ac{dizet} decoder, where each batch contains $\numMessages$ polynomials. During the training process, we train under a fixed $E_{\rm b}/N_{0}$, which is used to compute the noise variance $\sigma^2$. Furthermore, an exponential learning rate scheduler is used to gradually reduce the learning rate $\beta$ throughout $n_{\rm epoch}$ epochs. The steps for generating the $\messageIndex$th training example in a batch are as follows:
    \begin{enumerate}
        \item uniformly sample a message sequence $\messageSeq\jthExample$ from the set of all possible messages 
        $\{\messageSeq_0, \messageSeq_1, \dots, \messageSeq_{2^K-1}\}$;   
        \item map the message $\messageSeq\jthExample$ to the corresponding sequence of zeros $\zeroVector\jthExample = [\zero_0\jthExample,\zero_1\jthExample,\hdots,\zero_{\numZeros-1}\jthExample]$;
        \item compute the coefficients of a polynomial with the roots in $\zeroVector\jthExample$, i.e.,
        $\polySeqTX\jthExample=[x_0\jthExample,x_1\jthExample,\hdots,x_\numZeros\jthExample]$, and normalize $||\polySeqTX||_2^2$ to $\numZeros+1$;
        \item perturb $\polySeqTX\jthExample$ with \ac{AWGN} to yield the received sequence $\polySeqRX\jthExample$, where $y_k\jthExample = x_k\jthExample + w_k\jthExample$ and, $\forall k = 0, 1, \dots, K$, $w_k\jthExample \sim \complexnormal(0, \sigma^2)$.
    \end{enumerate}

    Next, we examine the \ac{NN}-based decoder, whose structure is motivated by other neural decoders in the literature \cite{decodingpolarcodes}. Table~\ref{tab:mlp_architecture} details its architecture, with $\hiddenlayers$ representing the number of neurons per hidden layer. To aid the training process, we utilize LeakyReLU activation functions in each hidden layer and incorporate dropout layers. The training procedure occurs in two stages. During the first stage, we learn $\lparametersMLP$ at higher $E_{\rm b}/N_{0}$ to avoid a suboptimal $\radius$, which can induce an error floor in \ac{BER} and \ac{BLER}. In the second stage, $\radius$ and $\phaseParameters$ are held constant, and $\mlpParameters$ are trained at lower $E_{\rm b}/N_{0}$ to improve robustness against noise. To generate the $j$th training example in a batch, we follow the \ac{dizet} procedure above, then compute the zero estimates $\bzerosestimate\jthExample$ using the companion matrix in \eqref{eq:companion_matrix}. An exponential learning rate scheduler is used in both stages, and the noise variance $\sigma^2$ is computed based on the associated $E_{\rm b}/N_{0}$.

    \subsection{Complexity Analysis}

    The \ac{dizet} decoder works by evaluating the received polynomial $\Y(z)$ at the $2\numZeros$ zeros in the constellation. Since evaluating $\Y(z)$ at a single zero takes on the order of $\numZeros$ multiplications, the complexity of the \ac{dizet} decoder is $\complexity(\numZeros^2)$. The proposed \ac{DL}-based decoder introduces two sources of complexity: the root-finding algorithm and the \ac{NN}. In this work, we perform root finding by computing the eigenvalues of a companion matrix, a computationally intensive method with complexity $\complexity(\numZeros^3)$. However, it is worth noting that certain root-finding algorithms, such as the Aberth method~\cite{aberth1973iteration}, can achieve a lower complexity of $\complexity(\numZeros^2)$. For the \ac{NN}, both the input and output size scale linearly with $\numZeros$, whereas the complexity of the hidden layers scales quadratically with $\hiddenlayers$. Since $\hiddenlayers \gg \numZeros$ for our experiments, the \ac{NN} complexity is on the order of $\complexity(\hiddenlayers^2)$. Nevertheless, the performance bottleneck in our implementation remains the eigenvalue computation, which dominates due to its cubic complexity.

	\begin{table}[t]
		\centering
		\caption{NN Decoder Architecture}
		\begin{tabular}{clc}
			\toprule
			\textbf{Layer \#} & \textbf{Layer description} & \textbf{Output} \\ 
			\midrule
			1 & Input Layer & \(2\numZeros\) \\ 
			2 & Dense + LeakyReLU & \(\hiddenlayers\) \\ 
			3 & Dropout (rate = 0.25) & \(\hiddenlayers\) \\ 
			4 & Dense + LeakyReLU & \(\hiddenlayers\) \\ 
			5 & Dropout (rate = 0.25) & \(\hiddenlayers\) \\ 
			6 & Dense (Output Layer) & \(\numZeros\) \\ 
			\bottomrule
		\end{tabular}
		\label{tab:mlp_architecture}
	\end{table}

    \section{Numerical results} \label{sec:results}

    In this study, we train models with $\numZeros=4,7,10$. For the \ac{dizet}-based approach, training comprises $n_{\rm epoch}=3 \times10^{4}$ epochs, where each batch contains $\numMessages=256$ polynomials. Training is conducted at $E_{\rm b}/N_{0}=10$~dB, and the margin threshold in \eqref{eq:loss_dizet} is set to $\threshold=1$. The initial learning rate is $\beta_{\rm i}=1\times10^{-2}$ and decays exponentially to $\beta_{\rm f}=1\times10^{-4}$ over the course of training. For the \ac{NN}-based approach, both stages of training comprise $n_{\rm epoch}=1.5 \times10^{4}$ epochs, where the learning rate decays exponentially from $\beta_{\rm i}$ to $\beta_{\rm f}$. In the first stage, training is conducted at $E_{\rm b}/N_{0}=10$~dB, while the second stage is trained at $E_{\rm b}/N_{0}=5$~dB. The number of neurons in the hidden layers of each \ac{NN}-based decoder is $\hiddenlayers=500, \: 1000, \: \text{and} \: 1500$ for $\numZeros=4, \: 7, \: \text{and} \: 10$, respectively.  
    
    \subsection{Comparison of Zero Patterns}

\begin{figure}[t]
        \centering
        \includegraphics[width=2.75in]{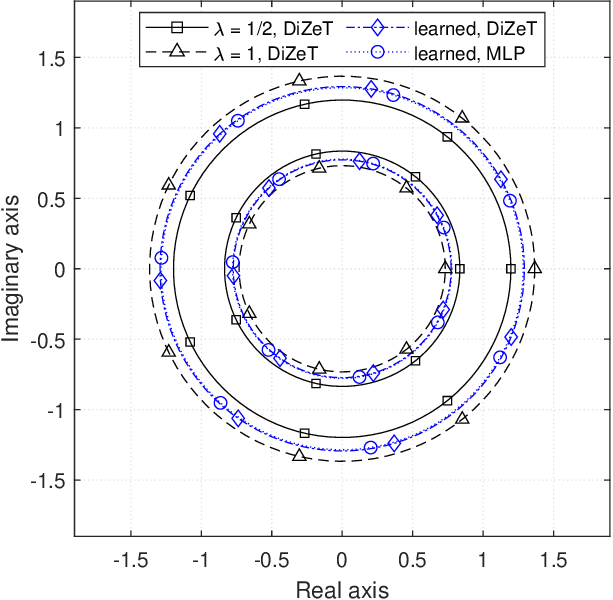}
        \caption{Zero constellations for $\numZeros=7$, showing the learned zero constellations together with Huffman \ac{BMOCZ} for $\lambda=1/2,1$.}	
        \label{fig:constellation_diagram}
    \end{figure}

    Fig.~\ref{fig:constellation_diagram} shows the zero constellations of various \ac{BMOCZ} schemes with $\numZeros=7$. The figure includes the constellation for Huffman \ac{BMOCZ} using the radius defined in \eqref{eq:dizet_radius} with $\lambda=1$ and $\lambda=1/2$. Additionally, the figure shows the constellations learned from Section~\ref{subsec:enc} and Section~\ref{subsec:dec} for the \ac{dizet} and \ac{NN}-based decoders, respectively. Notice that the learned phase parameters, $\phaseParameters$, are uniform in all cases. Furthermore, the learned radius, $\radius$, is similar for the two learned constellations and falls between the radii for $\lambda=1$ and $\lambda=1/2$. Table~\ref{tab:radii} displays the final radius for each scheme with $\numZeros=4,7,10$. As $\numZeros$ increases, we observe that the learned radii approach the value defined in \eqref{eq:dizet_radius} for $\lambda=1/2$. 

    \begin{table}[t]
        \caption{Radius of different schemes for $\numZeros=4,7,10$}
        \centering
        \label{tab:radii}
        \begin{tabular}{llccc}
            \toprule
            & & \multicolumn{3}{c}{\textbf{Radius}}\\\cline{3-5}
            \rule{0pt}{2.5ex}\textbf{Zero constellation} & \textbf{Decoder} & $\numZeros=4$ & $\numZeros=7$ & $\numZeros=10$\\
            \midrule
            \(\lambda = 1/2\) & DiZeT & \(1.3066\) & \(1.1974\) & \(1.1441\) \\ 
            \(\lambda = 1\) & DiZeT & \(1.5538\) & \(1.3667\) & \(1.2720\) \\ 
            learned & DiZeT & \(1.5457\) & \(1.2932\) & \(1.1822\) \\ 
            learned & NN & \(1.5831\) & \(1.2851\) & \(1.1963\) \\ 
            \bottomrule
        \end{tabular}
    \end{table}

    \subsection{Bit and Block Error Rate Performance}

    \begin{figure}[t]
		\centering
		\subfloat[Bit error rate performance.]
        {\includegraphics[width=3in]{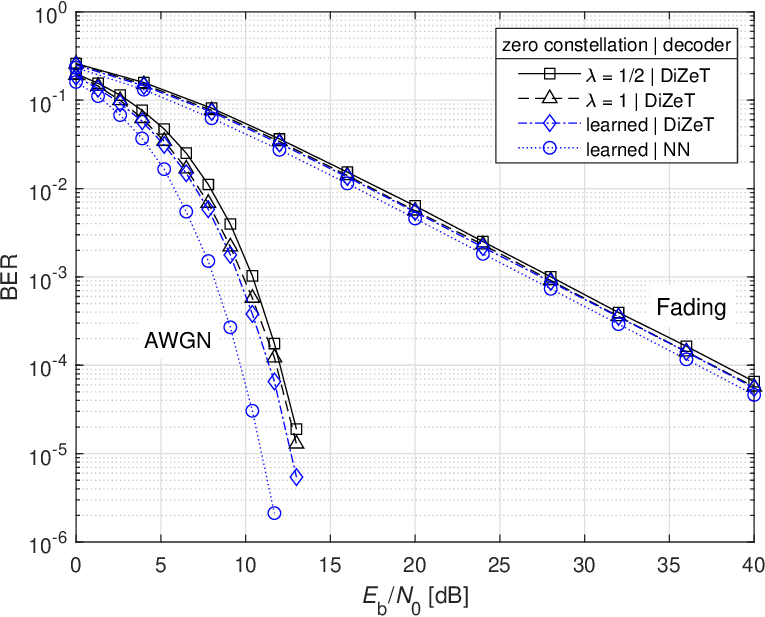}\label{subfig:ber}}~\\
		\subfloat[Block error rate performance.]{\includegraphics[width=3in]{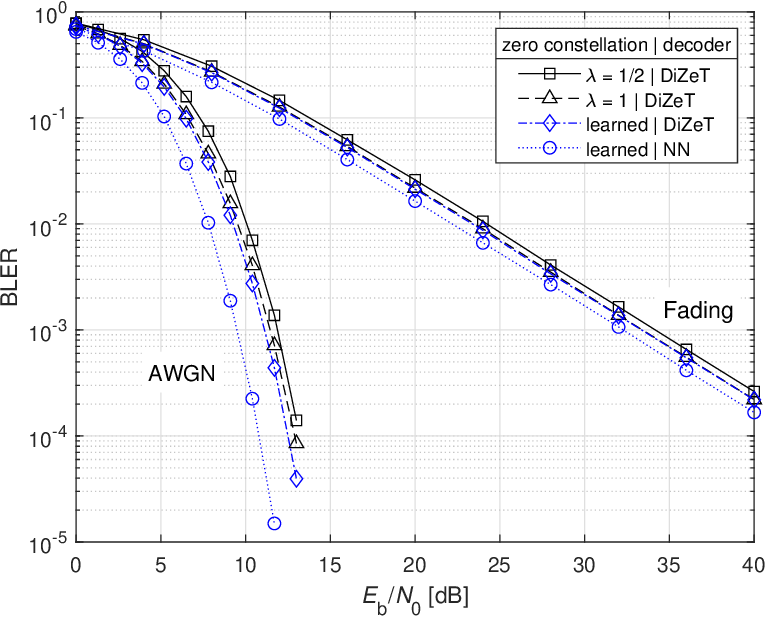}\label{subfig:bler}}	
		\caption{Performance of different schemes for $\numZeros=7$ in both \ac{AWGN} and flat-fading channels.}	
		\label{fig:awgn_error_curves}
	\end{figure}
    
   Here we consider the canonical \ac{BMOCZ} zero constellation (i.e., Huffman \ac{BMOCZ} with $\lambda=1/2,1$) as a benchmark and compare its performance to the learned zero constellations for the \ac{dizet} and \ac{NN}-based decoders with $\numZeros=4,7,10$. All schemes employ \ac{BMOCZ}-based \ac{OFDM} and frequency mapping, which utilizes $\numZeros+1$ active subcarriers, i.e., one subcarrier for each polynomial coefficient. With this approach, each \ac{OFDM} symbol observes flat-fading, so long as the bandwidth of the signal is less than the coherence bandwidth of the channel~\cite{huggins2024optimal}. In our simulations, we fix the \ac{IDFT} size to $32$~samples. In the \ac{AWGN} channel, each \ac{OFDM} symbol is perturbed by a noise sequence $\noiseSeq$ with $w_i\sim\complexnormal(0,N_0)$, where $N_0$ is the noise variance. In the fading channel, the $i$th \ac{OFDM} symbol is multiplied by $h_i\sim\complexnormal(0,1)$ and then perturbed by \ac{AWGN}. At every iteration through the simulations, we sample messages $\messageSeq\in\{0,1\}^\numZeros$ from a uniform distribution and normalize the energy of each transmitted sequence by setting $||\polySeqTX||_2^2=\numZeros+1$. We assume perfect synchronization at the receiver but no knowledge of the channel.

    Fig.~\ref{fig:awgn_error_curves} shows the \ac{BER} and \ac{BLER} curves simulated under \ac{AWGN} for each scheme with $\numZeros=7$. Observe that the \ac{NN}-based decoder achieves a large performance gain in both \ac{BER} and \ac{BLER}. Furthermore, the learned constellation for the \ac{dizet} decoder slightly outperforms the canconical \ac{BMOCZ} zero constellation using the radius in \eqref{eq:dizet_radius} with $\lambda=1$ and $\lambda=1/2$. Fig.~\ref{fig:awgn_error_curves} also shows the \ac{BER} and \ac{BLER} curves simulated in a flat-fading channel for each scheme with $\numZeros=7$. Although the difference is less pronounced, again the \ac{NN}-based decoder achieves a performance gain in both \ac{BER} and \ac{BLER}, with the BLER gain exceeding $1$ dB compared to all other schemes.  Moreover, the learned zero constellation for the \ac{dizet} decoder achieves comparable performance to \eqref{eq:dizet_radius} with $\lambda=1$ while slightly outperforming \eqref{eq:dizet_radius} with $\lambda=1/2$. 

\begin{figure}[t]
    	\centering
    	\includegraphics[width=3.0in]{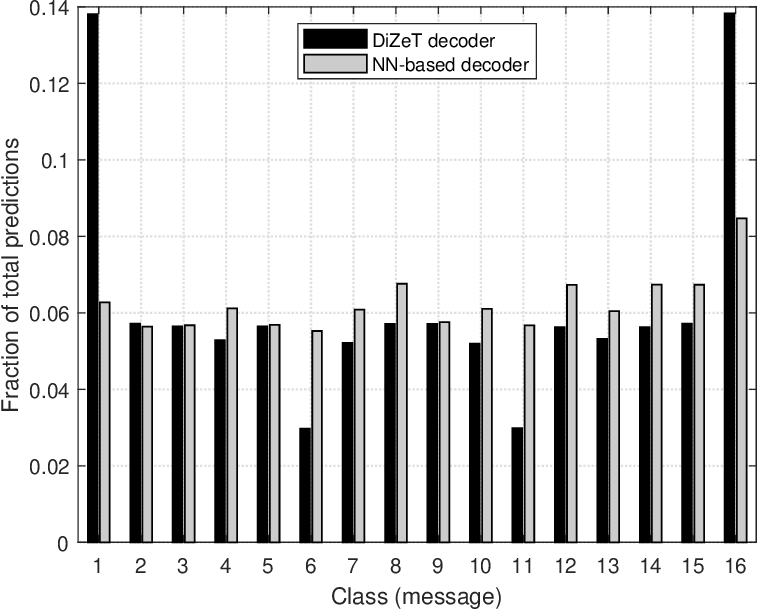}
    	\caption{Decoded messages for the \ac{dizet} and \ac{NN}-based decoders with $\numZeros=4$. For each class $50,000$ messages were decoded at $\ebno=-5$~dB. Classes one and sixteen correspond to the all zeros and all ones messages, respectively.}	
    	\label{fig:predictions_bar}
    \end{figure}

    To quantify the performance gain of our proposed schemes, we measure the $\ebno$ at which the associated error curves achieve a \ac{BLER} of $10^{-3}$. For each scheme, the relative gain in dB is computed as the absolute difference between the $\ebno$ measured for the worst performing scheme and that of all other approaches. These results are reported in Table~\ref{tab:results}, which displays the relative gain of each scheme in both \ac{AWGN} and flat-fading channels for $\numZeros=4,7,10$. Consistent with Fig.~\ref{fig:awgn_error_curves}(b), the \ac{NN}-based scheme achieves the largest performance gain in \ac{BLER} across all values of $\numZeros$. Furthermore, the learned constellations for the \ac{dizet} decoder achieve the second largest performance gains, although the gains are significantly smaller than those achieved by the \ac{NN}, particularly as $\numZeros$ increases. The gain of the \ac{NN}-based decoders gain can be partially attributed to their lack of bias compared to the \ac{dizet} decoder, e.g., see Fig.~\ref{fig:predictions_bar}. Specifically, the \ac{dizet} decoder estimates the transmitted message as all zeros or all ones a disproportionate amount of time, a behavior not observed with the \ac{NN}-based decoder. 
 
    \begin{table}[t]
        \caption{Relative gain of different schemes for $\numZeros=4,7,10$}
        \centering
        \setlength{\tabcolsep}{2.24pt}
        \label{tab:results}
        \begin{tabular}{*{2}{l}@{\hskip 5pt}*{3}{c}@{\hskip 5pt}*{3}{c}}
            \toprule
             & & \multicolumn{3}{@{\hskip 5pt}c@{\hskip 5pt}}{\textbf{Gain (dB), AWGN}} & \multicolumn{3}{@{\hskip 5pt}c@{\hskip 5pt}}{\textbf{Gain (dB), fading}} \\
            \cmidrule(lr){3-5} \cmidrule(lr){6-8} 
            \rule{0pt}{2.5ex}\raggedright\shortstack[l]{\textbf{Zero} \\ \textbf{const.}} & \textbf{Decoder} & $\numZeros=4$ & $\numZeros=7$ & $\numZeros=10$ & $\numZeros=4$ & $\numZeros=7$ & $\numZeros=10$ \\
            \midrule
            \(\lambda = 1/2\) & DiZeT & \(0.00\) & \(0.00\) & \(0.00\) & \(0.00\) & \(0.00\) & \(0.00\)\\
            \(\lambda = 1\) & DiZeT & \(1.27\) & \(0.50\) & \(0.20\) & \(1.62\) & \(0.78\) & \(0.24\)\\
            learned & DiZeT & \(1.24\) & \(0.71\) & \(0.37\) & \(1.54\) & \(0.83\) & \(0.46\)\\
            learned & NN & \(3.66\) & \(2.30\) & \(1.78\) & \(2.92\) & \(2.20\) & \(1.57\)\\
            \bottomrule
        \end{tabular}
    \end{table}  
    
    \section{Concluding Remarks} \label{sec:conclusion}

    In this study, we propose \ac{ML} frameworks to design zero constellations for \ac{BMOCZ}. We first consider \ac{BMOCZ} implemented with a \ac{dizet} decoder and learn the zero constellation parameters to minimize \ac{BER}. Next, we introduce an \ac{NN}-based decoder for \ac{BMOCZ} that utilizes the zeros of the received polynomial for decoding. After jointly learning the decoder and zero constellation parameters under \ac{AWGN}, we show that the \ac{NN}-based decoder can generalize to flat-fading channels without additional inputs or \ac{CSI}. Finally, we show that the learned zero constellations outperform the canonical, Huffman \ac{BMOCZ} zero constellation, particularly under \ac{AWGN}.
    
    The results presented in this work suggest that \ac{ML}-based optimization for \ac{BMOCZ} is a promising approach to address the challenge of training in a fading channel. In future work, we will focus on refining the \ac{NN}-based decoder to improve performance and reduce complexity, with additional studies to emphasize its applicability to higher-order \ac{MOCZ} schemes and longer sequence lengths. Additionally, we will consider learning zero constellations that are robust against ubiquitous hardware impairments, such as timing and carrier frequency offsets, to further improve the practicality of \ac{MOCZ}. 
	
    \bibliographystyle{IEEEtran}
    \bibliography{references}
	
\end{document}